\documentclass{optica-article}

\journal{opticajournal} 

\articletype{Research Article}

\usepackage{lineno}

\usepackage{amsmath} 
\usepackage[utf8]{inputenc}
\usepackage{xcolor}

\begin{document}
	
	\title{Unveiling the origins of quasi-phase matching spectral imperfections in thin-film lithium niobate frequency doublers}

	\author{Jie Zhao,\authormark{1,3,*} Xiaoting Li,\authormark{2,3} Ting-Chen Hu,\authormark{1} Ayed Al Sayem,\authormark{1} Haochuan Li,\authormark{2} Al Tate,\authormark{1} Kwangwoong Kim,\authormark{1} Rose Kopf,\authormark{1} Pouria Sanjari,\authormark{1,4} Mark Earnshaw,\authormark{1} Nicolas K. Fontaine,\authormark{1} Cheng Wang,\authormark{2} and Andrea Blanco-Redondo\authormark{1,5}}
	
	\address{\authormark{1} Nokia Bell Labs, 600 Mountain Avenue, Murray Hill, NJ 07974, USA\\
		\authormark{2} Department of Electrical Engineering \& State Key Laboratory of Terahertz and Millimeter Waves, City University of Hong Kong, Hong Kong, China\\
		\authormark{3} These authors contributed equally to this work.\\
		\authormark{4} Department of Electrical \& Systems Engineering, University of Pennsylvania, Philadelphia, PA 19104, USA\\
		\authormark{5}{CREOL, The College of Optics and Photonics, University of Central Florida, Orlando, FL 32816, USA} }
	\email{\authormark{*}jie.2.zhao@nokia-bell-labs.com} 
	
	
	\begin{abstract*} 
		Thin-film lithium niobate (TFLN) based frequency doublers have been widely recognized as essential components for both classical and quantum optical communications. Nonetheless, the efficiency of these devices is hindered by imperfections present in the quasi-phase matching (QPM) spectrum. In this study, we present a thorough analysis of the spectral imperfections in TFLN frequency doublers with varying lengths, ranging from 5~mm to 15~mm. Employing a non-destructive diagnostic method based on scattered light imaging, we identify the sources and waveguide sections that contribute to the imperfections in the QPM spectrum. Furthermore, by mapping the TFLN film thickness across the entire waveguiding regions, we successfully reproduce the QPM spectra numerically, thus confirming the prominent influence of film thickness variations on the observed spectral imperfections. This comprehensive investigation provides valuable insights into the identification and mitigation of spectral imperfections in TFLN-based frequency doublers, paving the way toward the realization of nonlinear optical devices with enhanced efficiency and improved spectral fidelity.
	\end{abstract*}
	
	\section{Introduction}
	Thin-film periodically poled lithium niobate (PPLN) waveguides offer a compelling platform for achieving highly efficient wavelength conversion devices, leveraging their high second-order nonlinear coefficient ($\chi^{(2)}$) and tight confinement of optical modes. These waveguides have found diverse applications in second-harmonic generation (SHG)~\cite{wang2018g,Lu2019c}, entangled photon-pair generation~\cite{zhao2020a,ma2020}, optical parametric amplification~\cite{ledezma2022}, optical isolation~\cite{Abdelsalam2020}, and all-optical switching~\cite{Guo2021}. However, compared to their bulk counterparts, these waveguides exhibit an increased susceptibility to fabrication inhomogeneities, presenting a significant challenge to their overall performance~\cite{zhao2020c,kuo2022,Tian2021a}. These inhomogeneities can give rise to unfavorable effects in the QPM spectrum, including broadened central peaks and unwanted side lobes, resulting in reduced conversion efficiency. Furthermore, as phase errors accumulate along the waveguide's length, longer devices experience more pronounced impacts from fabrication non-uniformity, hindering the further enhancement of power conversion efficiency~\cite{Santandrea2019c}. 
	To date, the demonstrated thin-film PPLN devices have typically had lengths ranging between 4 and 6~mm. As a result, the overall conversion efficiency of these devices remains lower than what has been reported for their bulk counterparts~\cite{Parameswaran2002}. Therefore, it is imperative to direct research efforts towards investigating approaches that can enable the fabrication of longer thin-film PPLN devices, while ensuring good spectral fidelity. This research is crucial to bridge the performance gap with their bulk counterparts and fully unlock the potential of thin-film PPLN technology. 
	
	Prominent studies have examined the impact of fabrication inhomogeneities on the QPM spectrum from Ti-indiffused and diced Zinc-indiffused bulk lithium niobate (LN) waveguides~\cite{Santandrea2019b,santandrea2019,Gray2020}, with a specific focus on waveguide width variation as the primary error source. In particular, in Ref.~\cite{Santandrea2019b}, 83~mm Ti-indiffused PPLN waveguides were diced into shorter sections to analyze the generated QPM spectra from different regions of the waveguides, thereby revealing the evolution of the QPM spectrum along the waveguide's length. In the context of thin-film PPLN devices, numerical methods have been employed to replicate the measured QPM spectra based on an estimated TFLN film thickness profile~\cite{Tian2021a,Xue2022,kuo2022}. However, comprehensive experimental investigations akin to those described in Ref.~\cite{Santandrea2019b}, which would greatly enhance our understanding of QPM spectral imperfections and benefit the development of longer devices, are currently lacking for thin-film PPLN devices.
	
	Here, we introduce a non-destructive optical diagnostic method that enables the visualization of the QPM spectrum at any location along the thin-film PPLN waveguide. To accomplish this, we utilize a monochrome camera positioned perpendicular to the chip surface to capture the scattered second-harmonic (SH) light across the entire PPLN waveguiding region. By acquiring multiple images while sweeping the pump wavelengths, the local QPM spectra at different locations on the waveguide can subsequently be calculated and obtained. This innovative technique uncovers the contributions from different sections of the waveguide to the final QPM spectrum, facilitating an in-depth understanding of the imperfections observed in the measured spectra. To evaluate the efficacy of our approach, we conducted investigations on thin-film PPLN waveguides with lengths of 5~mm, 7.5~mm, 12.5~mm, and 15~mm. Our findings indicate that the observed imperfections in the spectra primarily stem from variations in TFLN film thickness. Subsequently, we performed a thorough mapping of the film thickness across the entire waveguiding regions. This comprehensive mapping allowed us to successfully numerically reproduce the measured QPM spectra, further confirming the significant influence of film thickness variations on the spectral imperfections.
	
	\begin{figure}[htbp]
		\centering
		\includegraphics[width=0.95\textwidth]{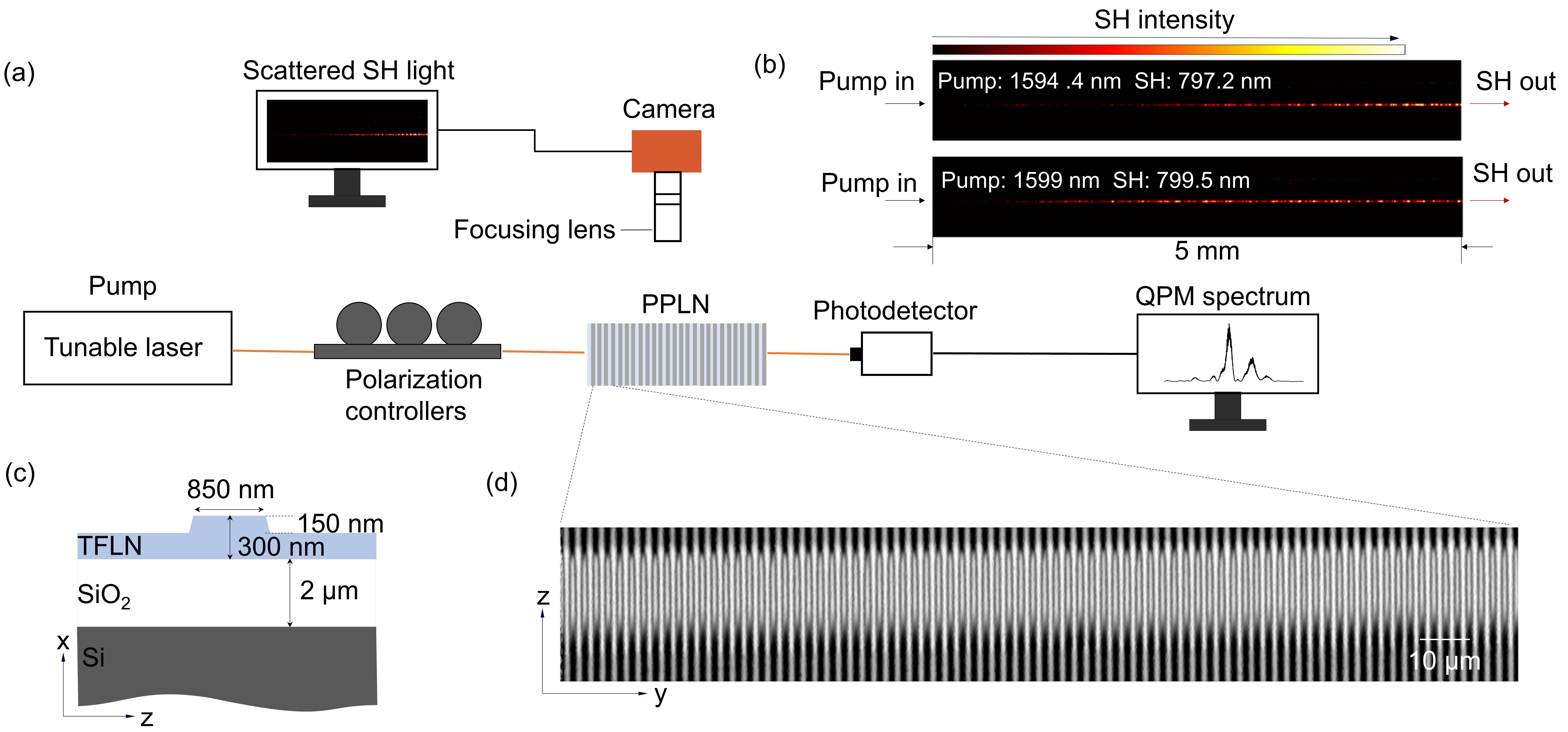}
		\caption{\label{fig:1} (a) Schematic of the experimental setup for SHG characterization. (b) Images of the collected scattered SH light from a 5~mm long PPLN waveguide with the pump wavelength at 1594.4~nm and 1599~nm respectively. (c) Cross-section of the PPLN waveguides. (d) Second-harmonic microscope image of the periodically poled lithium niobate thin film.} 
	\end{figure}
	
	\section{Methods}
	The thin-film PPLN waveguides were fabricated using 5 mol\% MgO-doped 300~nm x-cut lithium niobate on insulator (LNOI) wafers. In this report, all measured waveguides have a targeted etching depth of 150~nm and a waveguide top width of approximately 850~nm, as shown in Fig.~1(c). These waveguides were designed for efficient SHG from telecommunication wavelengths to near-visible wavelengths. The required poling period for QPM is about 2.46~$\mu$m, and all the interacting waves are in the fundamental TE mode. 
	
	In the initial fabrication step, poling electrodes with lengths varying from 5~mm to 15~mm were formed on the surface of TFLN using photolithography, while maintaining a fixed poling period of 2.46~$\mu$m for each electrode. High-voltage pulses, as described in Ref.~\cite{Zhao2020}, were then applied to these electrodes for periodic poling of the TFLN. Figure~1(d) presents a representative second-harmonic (SH) microscope image of the poled area, revealing domain structures with ideal uniformity and duty cycles, which are crucial for efficient frequency conversion. Multiple SH microscope images were generated at different locations along each waveguide. It is worth emphasizing that achieving high-fidelity poling of TFLN with long lengths and small periods has been challenging. As the poling periods decrease, adjacent domains tend to merge, while longer waveguides introduce more variations in the duty cycle. Despite these challenges, we have achieved remarkable results in our devices. To the best of our knowledge, the 7.5~mm long PPLN waveguide we have fabricated is currently the longest demonstrated on the x-cut TFLN platform. Here, even for the 15~mm long poling electrodes, the calculated poling duty cycle remains highly uniform, at 52.83\% $\pm$ 2.28\%. Therefore, contributions from imperfect periodic poling of TFLN (as discussed in detail through theoretical calculations in Ref.~\cite{Fejer1992a}) to the QPM spectra are not considered in our subsequent discussions and numerical simulations. The waveguides were fabricated in the poled areas through aligned electron beam lithography and dry etching, followed by a wet etching process for sidewall deposition cleaning. Finally, the chip edges were cleaved to ensure optimal edge coupling.
	
	The experimental setup for SHG characterization is depicted in Fig.~1(a). A tunable laser operating at telecommunication wavelengths serves as the pump source for the PPLN waveguides, with its polarization adjusted by a set of polarization controllers to ensure TE mode injection at the waveguide's input facet. Light is coupled into and out of the chip using tapered lensed fibers to optimize the coupling efficiency. While sweeping the pump wavelength, the corresponding SH light is collected at the output waveguide facet using a lensed fiber and detected by a Si photodetector, following the conventional method for obtaining the QPM spectrum. Additionally, we developed and employed a novel approach for SHG characterization, which not only provides QPM spectrum information at any location along the waveguide but also facilitates the identification of sources causing spectral imperfections. Specifically, while sweeping the pump wavelengths, a monochrome camera (Allied Vision Alvium 1800 U-511) was used to image the scattered SH light along the waveguide region. For example, Fig.~1(b) presents the measured scattered SH light images from a 5~mm thin-film PPLN waveguide with pump wavelengths of 1594.4~nm and 1599~nm, respectively. As we can see in these images, SH light with different wavelengths is scattered from distinct parts of the waveguide. In the case of this waveguide, the 799.5 nm SH light forms an unwanted side lobe in the QPM spectrum, and Fig.~1(b) indicates that it is primarily generated near the initial section of the PPLN. Such information proves valuable for comprehending and addressing QPM spectral imperfections, particularly in longer waveguides, as further elaborated in the subsequent sections.
	
	\section{Results}
	We measured SHG from waveguides with lengths of 5~mm, 7.5~mm, 12.5~mm, and 15~mm. Note that these waveguides are located on the same chip, and their relative positions are depicted in Fig.~3(a). The QPM spectra obtained by the conventional method (as described in the previous paragraph) are shown as the blue curves in Figs.~2(c), (f), (i), and (l). Upon comparing these plots, it becomes evident that increasing the waveguide length leads to the generation of more side lobes, thereby diminishing the spectrum fidelity and the overall achievable conversion efficiency. The variations in the main peak wavelengths can be attributed to a combination of slight differences in the designed waveguide widths and the film thicknesses from one waveguide to another, as illustrated in Fig.~3(b). The spectrum fidelity values, calculated using the definition from Ref.~\cite{Santandrea2019c}, are 0.46, 0.24, 0.17, and 0.15 respectively for the four lengths. To gain deeper insights into the origins of these side lobes, we captured scattered SH light images for each waveguide while sweeping the pump wavelengths, similar to the representative images shown in Fig.~1(b). By integrating the pixel values in each column of these images (perpendicular to the waveguide's direction), we obtained mappings of the SH light in terms of propagation length and pump wavelength, which are presented as color images in Figs.~2(a), (d), (g), and (j). These visual representations conveniently illustrate the evolution of the QPM spectrum along the waveguide.
	Moreover, end-point QPM spectra, which refer to the QPM spectra obtained at the end of the PPLN waveguides, can be generated by averaging the final column values of the SH light mappings. They are depicted by the yellow curves in Figs.~2(c), (f), (i), and (l). The close alignment between these curves and the QPM spectra obtained through the conventional method provides compelling evidence that valid QPM spectra can be obtained using the scattered light imaging technique at any location along the PPLN waveguide. Therefore, we analyze the imperfections in the QPM spectra based on the SH light mappings. Notably, these mappings exhibit a consistent trend: side peaks with longer wavelengths than the main peaks predominantly arise from the initial section of the waveguides, while the SH light from the main peaks only begins to emerge at approximately 2-3~mm away from the waveguides' starting points. This similarity suggests the presence of one or several waveguide geometry parameters that consistently vary along the direction of light propagation across the entire chip. 
	
	\begin{figure}[htbp]
		\centering
		\includegraphics[width=0.95\textwidth]{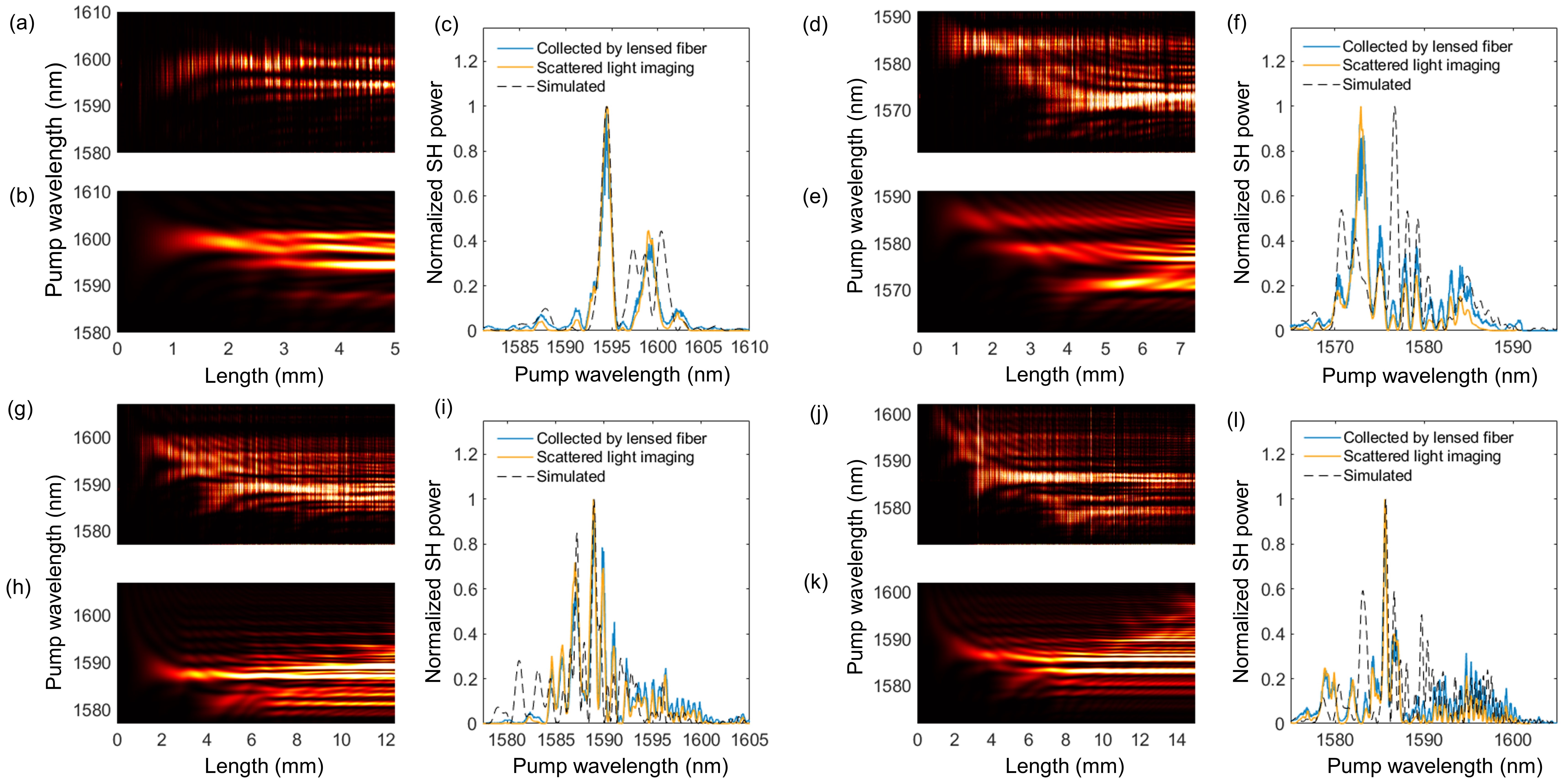}
		\caption{\label{fig:2} Measured and simulated QPM spectrum mappings and end-point QPM spectra from thin-film PPLN waveguides with different lengths of 5 mm (a)-(c), 7.5 mm (d)-(f), 12.5 mm (g)-(i), and 15 mm (j)-(l). (a), (d), (g), and (j): Measured QPM spectrum mappings using scattered-light imaging. (b), (e), (h), and (k): Calculated QPM spectrum mappings based on measured TFLN film thickness profiles. (c), (f), (i), and (l): Measured end-point QPM spectra using the conventional method (blue curves), as well as scattered-light imaging (yellow curves), overlaid with calculated plots (dashed black curves) based on TFLN film thickness profiles.} 
	\end{figure}
	
	For uniform waveguides, the QPM spectrum ($\Phi$, normalized per unit length) can be calculated as~\cite{Tian2021a,Santandrea2019c}: 
	\begin{equation}
		\begin{aligned} 
			\Phi = \text{sinc}(\frac{\Delta\beta L}{2})\text{exp}(i\frac{\Delta\beta L}{2}), \\
			\Delta\beta = \frac{2\pi}{\lambda_{\text{SH}}}n_{\text{SH}} - 2\frac{2\pi}{\lambda_{\text{pump}}}n_{\text{pump}} - \frac{2\pi}{\Lambda},
		\end{aligned} 
	\end{equation}
	where $L$ is the length of the waveguide, $\lambda_\text{pump, SH}$ denotes the pump and SH wavelengths, and $n_{\text{pump, SH}}$ indicates the effective mode indices for the pump and SH light. To determine the type of fabrication errors causing these spectrum imperfections, we calculated the partial derivatives of the momentum mismatch ($\Delta\beta$, as defined in Eq.~1) with respect to the waveguide top width $w$, etching depth $h_{\text{etch}}$, and film thickness $t_{\text{LN}}$ as follows: 
	\begin{equation}
		\begin{aligned} 
			(\frac{\partial \Delta\beta}{\partial w}) _{\{h_{\text{etch}}=154~nm,~t_{\text{LN}}=303~nm,~\lambda_{\text{pump}}=1580~nm\}} &= - 0.485~\mu m^{-2}, \\ 
			(\frac{\partial \Delta\beta}{\partial h_{\text{etch}}}) _{\{w=850~nm,~t_{\text{LN}}=303~nm,~\lambda_{\text{pump}}=1580~nm\}} &= 1.842~\mu m^{-2}, \\
			(\frac{\partial \Delta\beta}{\partial t_{\text{LN}}}) _{\{w=850~nm,~h_{\text{etch}}=154~nm,~\lambda_{\text{pump}}=1580~nm\}} &= -4.018~\mu m^{-2}. \\
		\end{aligned}
	\end{equation}
	The calculations reveal that $\Delta\beta$ of the waveguide structure employed in this study is most sensitive to variations in film thickness. In addition, based on the general framework developed in Ref.~\cite{Santandrea2019c}, we further estimated the contribution from each individual waveguide parameter to the QPM spectrum infidelity for all four waveguides, and the results are summarized in Table 1. This estimation assumes that the spectrum infidelity is primarily influenced by a single waveguide parameter, which exhibits 1/f noise to account for the long-range correlations arising from the fabrication processes~\cite{santandrea2019}. The similarity in the calculated variations for each parameter across all four waveguides provides further evidence of the consistent nature of the fabrication variations throughout the entire chip.
	\begin{table}[!h]
		\begin{center}
	\caption{\label{tab1} Estimated waveguide geometry variation for all four waveguides based on the measured QPM spectra.}
			\begin{tabular}{||c c c c||} 
				\hline
				Waveguide length & $\Delta w$ & $\Delta h_{\text{etch}}$ & $\Delta t_{\text{LN}}$ \\ [0.5ex] 
				\hline\hline
				5~ mm & 13.16~nm & 3.47~nm & 1.59~nm \\ 
				\hline
				7.5~mm & 22.04~nm & 5.81~nm & 2.66~nm \\
				\hline
				12.5~mm & 20.70~nm & 5.46~nm & 2.50~nm \\
				\hline
				15~mm & 20.42~nm & 5.38~nm & 2.47~nm \\
				\hline
			\end{tabular}
		\end{center}
	\end{table}
	
	We conducted precise measurements using atomic force microscopy (AFM) on a twin chip to assess the variations in waveguide width and etching depth along a 15 mm long waveguide. The calculated errors for both parameters were found to be lower than the estimated values mentioned in Table 1. On the other hand, the estimated variations in film thickness appear more plausible, as such levels of long-range variation can often arise from the chemical-mechanical polishing process and are consistent with the labeled thickness uniformity value provided by the wafer vendor. In order to examine the film thickness variations, we used Filmetrics to measure the thickness of TFLN across the entire etched chip. The measurement was performed with a step size of 200 $\mu$m, and the resulting mapping is illustrated in Fig.~3(a). It is worth noting that this measurement was conducted subsequent to the dry-etching process of the waveguide. Therefore, the plotted thickness values in Fig.~3 represent the raw measured data, augmented by a constant etching depth of 154~nm. As depicted in Fig.~3(a), it is evident that the film thickness exhibits consistent variations across the entire chip, which aligns with our hypothesis based on the QPM spectrum mappings shown in Fig.~2. Specifically, the thickness gradually increases from the left edge of the chip (the input port of the PPLN) and subsequently decreases towards the right side (the output port of the PPLN). In addition to the overall mapping, Fig.~3(b) provides a detailed plot of the measured TFLN thickness along four specific waveguides with lengths of 5 mm (blue), 7.5 mm (red), 12.5 mm (yellow), and 15 mm (black). The dashed lines in the figure indicate the end of the PPLN regions. It is important to note that the chip analyzed in this study was obtained from an area near the edge of the wafer. The thickness profile, which exhibits a high-order polynomial trend, observed in this specific chip was not present in subsequent measurements conducted on chips extracted from regions in the center of the wafer.
	\begin{figure}[htbp]
		\centering
		\includegraphics[width=0.95\textwidth]{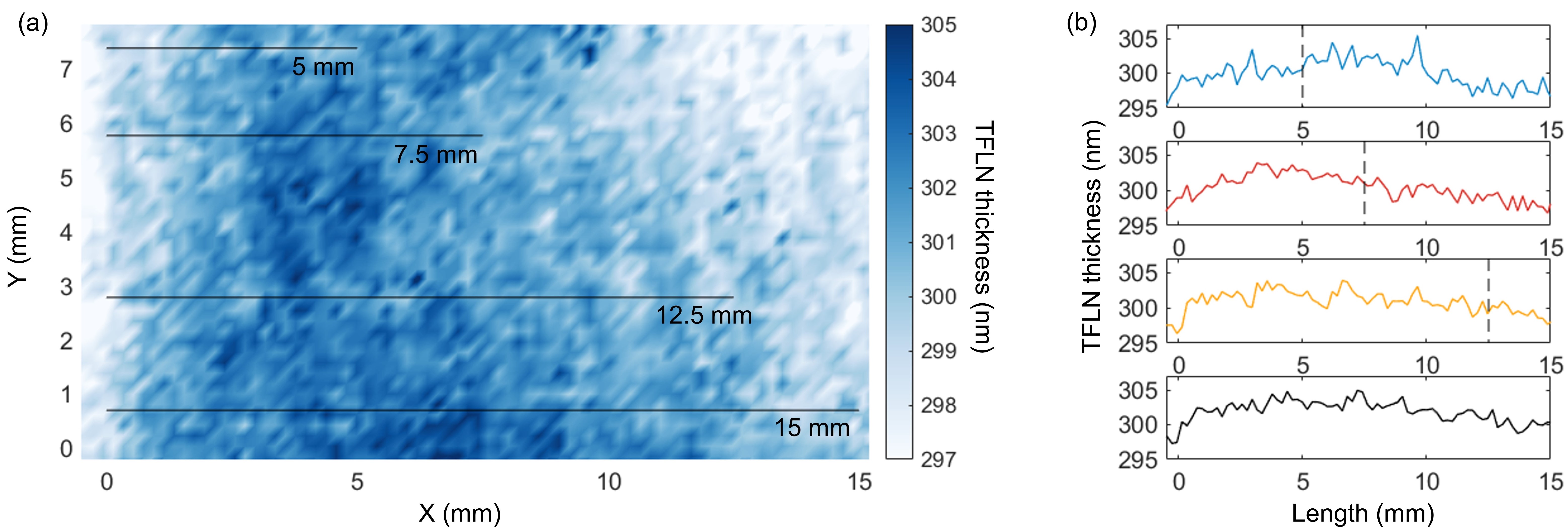}
		\caption{\label{fig:3} (a) Measured TFLN film thickness mapping over the entire chip, with the black lines showing the position of PPLN waveguides. (b) Measured TFLN thickness along the waveguides with lengths of 5~mm (blue), 7.5~mm (red), 12.5~mm (yellow), and 15~mm (black). The black dashed lines indicate the end of the PPLN region for each waveguide.} 
	\end{figure}
	
	Utilizing these film thickness values, we numerically calculated the QPM spectrum mapping for all four waveguides, based on the methods described in Refs.~\cite{Helmfrid1992,santandrea2019}. These mappings are presented in Figs.~2(b), (e), (h), and (k). We also depicted the end-point QPM spectra in Figs.~2(c), (f), (i), and (l), overlaid with the curves acquired experimentally.  
	A thorough comparison of these plots highlights the successful reproduction of the main side lobes observed in the measured QPM spectra for all four waveguides. Moreover, the simulations accurately captured the evolution trend evident in the measured QPM spectrum mappings. Several factors may contribute to the observed discrepancies between the measured and simulated mappings in this study. These factors include local variations in waveguide widths and etching depth, which are considered secondary factors here and thus are not included in the numerical calculations. Furthermore, considering the high sensitivity of the spectra to film thickness variations, even slight deviations from the true values, such as the absence of fine features in the captured thickness profile and errors resulting from imperfect measurement accuracy, can significantly contribute to the observed disparities. Taking the 15~mm PPLN waveguide as an example, in Fig.~4, we show that slight local variations in the film thickness profile can provide better alignment to the measured QPM spectra. The optimized film thickness profile, as shown in Fig.~4(a), was obtained using the Particle Swarm Optimization algorithm, aiming to reproduce the QPM spectra at various locations along the waveguide (Fig.~4(b)). Note that the difference between the optimized thickness profile and the raw data can be introduced by other waveguide parameters other than film thickness, such as waveguide width and etching depth. As for the 12.5~mm long waveguide, we could achieve a better matching with the measured QPM spectrum mapping (shown in Fig.~2(h)) by adding 2~nm to the poling period, shifting the whole spectrum to longer wavelengths, though the alignment with the end-point QPM spectrum would be compromised. It is important to emphasize that the goal of this work is not to perfectly replicate the measured QPM spectrum mapping, as achieving this level of precision with a single parameter variation (in this case, film thickness) is unrealistic. Nevertheless, the remarkable agreement observed between the simulated and measured QPM spectra provides strong evidence supporting the hypothesis that the observed imperfections primarily originate from variations in the film thickness along the waveguiding region. 
	
	\begin{figure}[htbp]
		\centering
		\includegraphics[width=0.95\textwidth]{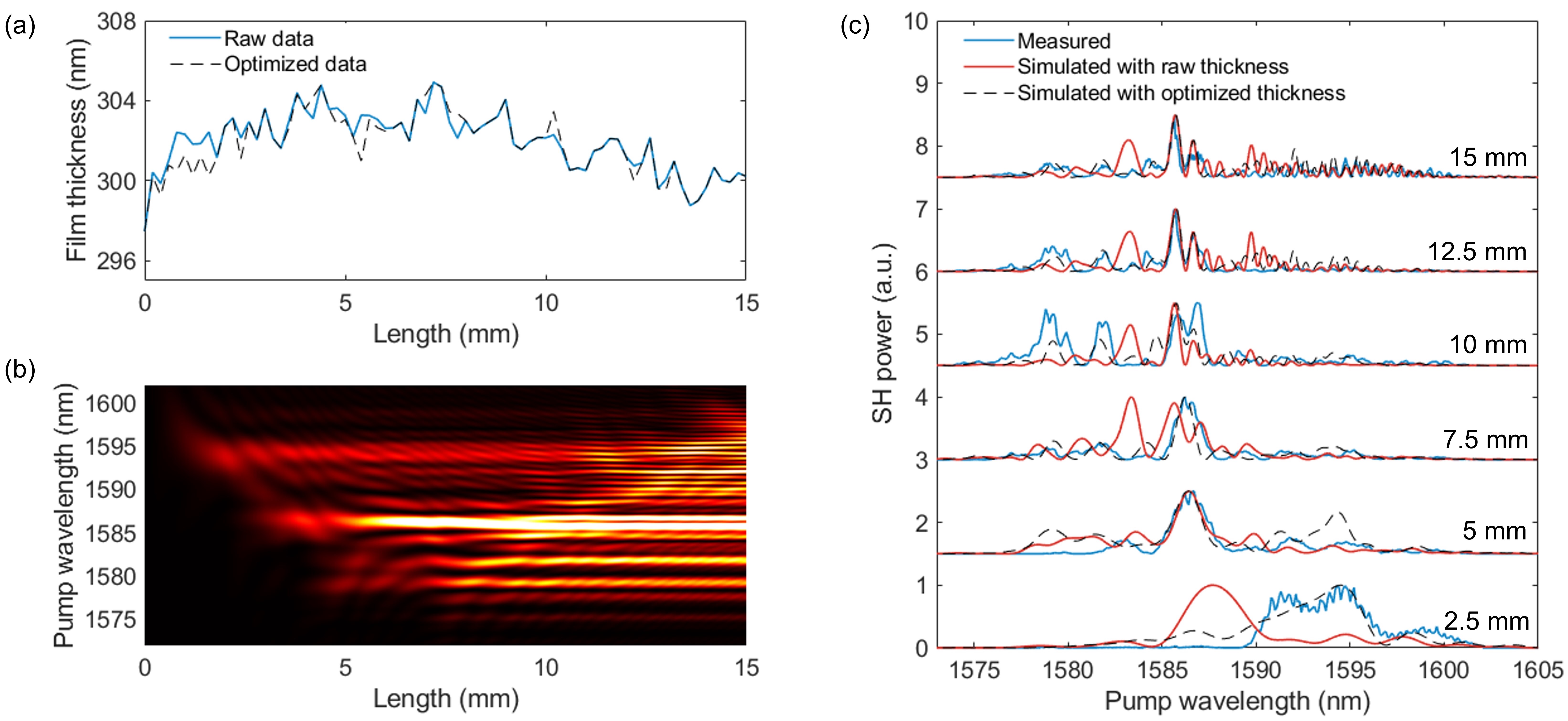}
		\caption{\label{fig:4} (a) Comparison of the measured and optimized film thickness profile for the 15~mm long waveguide. (b) Simulated QPM spectrum mapping using the optimized thickness profile. (c) QPM spectra at different locations along the waveguide obtained through the experiment (solid blue curve), simulations with raw thickness profile (solid red curve), as well as simulations with optimized thickness profile (dashed black curve).} 
	\end{figure}
	
	\section{Conclusion}
	In conclusion, our study presents a non-destructive optical diagnostic method for assessing imperfections in QPM spectra from thin-film PPLN waveguides. This method allows for the extraction of QPM spectra at any position along the waveguide without the need for physically dicing the waveguides. By comparing the measured spectrum mappings with ideal ones, we can identify specific regions that contribute to unwanted distortions in the QPM spectra, providing crucial insights into the sources of these spectral imperfections. We applied this approach to investigate the QPM spectrum imperfections in four fabricated thin-film PPLN waveguides with lengths of 5 mm, 7.5 mm, 12.5 mm, and 15 mm. Our analysis revealed a consistent trend in the QPM spectrum mappings across these waveguides. Combining our experimental results with numerical simulations, we identified variations in TFLN thickness along the waveguide as the primary source of the observed spectral imperfections. Moreover, we successfully replicated the main features of the measured QPM mappings using film thickness data obtained from Filmetrics measurements. The strong alignment between the simulated and measured mappings serves to validate our conclusions.
	
	There are several potential avenues for improving the fidelity of QPM spectra. One straightforward approach is to fabricate PPLN waveguides in an area close to the center of the LNOI wafer, where the film thickness is expected to be more uniform and exhibit a reduced variation trend observed in this study. Another promising strategy involves designing waveguide structures that are less sensitive to thickness variations, as explored in simulations detailed in Ref.~\cite{kuo2022}, which may come at the cost of reduced conversion efficiency. Additionally, obtaining a detailed TFLN thickness mapping prior to waveguide fabrication and compensating for thickness variations through adjustments in other waveguide parameters such as poling period and waveguide width is another viable option. It is also possible to design folded waveguides that possess a smaller area with more uniform film thickness~\cite{Weigel2018b}, although this would require additional components to control the accumulated phase mismatch in the curved waveguides.
	
	Taken together, our research provides new insights to the understanding of imperfections in QPM spectra originating from thin-film PPLN waveguides. The integration of experimental findings with numerical simulations sheds light on the path toward the development of long thin-film PPLN devices with enhanced overall conversion efficiency. These insights hold great potential for benefiting applications in both classical and quantum communication domains, opening up new possibilities for enhanced performance and functionality.

	
	\begin{backmatter}
		\bmsection{Funding} 
		Research Grants Council, University Grants Committee (CityU 11204820, N\_CityU113/20)
		
		\bmsection{Acknowledgments} 
		Jie Zhao would like to thank Haowen Ren, Yanqi Luo, and Yang Li for their valuable discussions and assistance in the fabrication process.
		
		\bmsection{Disclosures}
		The authors declare no conflicts of interest.

		\bmsection{Data Availability Statement}
		Data underlying the results presented in this paper are not publicly available at this time but may be obtained from the authors upon reasonable request.

	\end{backmatter}
	
	

\end{document}